\documentclass[a4paper,11pt]{article}
\usepackage{jinstpub} % jinstpub loads natbib internally

\usepackage{amsmath,amssymb}

\usepackage{graphicx}
\usepackage{subcaption}
\usepackage{booktabs}
\usepackage{float}
\DeclareUnicodeCharacter{2009}{ }
\DeclareUnicodeCharacter{03BC}{$\mu$}
\DeclareUnicodeCharacter{2212}{-}

\title{\boldmath GPU-based track-finding for the J-PARC muon g-2/EDM experiment}
\author[a]{Hridey Chetri\footnote{First author}}
\author[b]{Deepak Samuel\footnote{Corresponding author}}
\author[a]{Saurabh Sandilya}
\author[c]{Takashi Yamanaka}
\author[d]{Tsutomu Mibe}
\author[e]{Taikan Suehara}
\affiliation[a]{Indian Institute of Technology Hyderabad, \\Hyderabad, India}
\affiliation[b]{Central University of Karnataka,\\ Karnataka, India}
\affiliation[c]{Kyushu University,\\ Fukuoka, Japan}
\affiliation[d]{Institute of Particle and Nuclear Studies (IPNS),\\ KEK, Japan}
\affiliation[e]{ICEPP, University of Tokyo,\\ Tokyo, Japan}

\emailAdd{deepaksamuel@cuk.ac.in}

\abstract{The muon \textit{g-2}/EDM experiment at J-PARC is designed to precisely measure the muon's magnetic moment and electric dipole moment, driven by discrepancies between theory and previous experiments. One of the key challenges in the experiment is the fast reconstruction of positron tracks from multiple muon decays within a short time span causing an event pileup.
Results from simulation studies have shown expected results in terms of efficiency and accuracy of track reconstruction. However, the execution time for the entire analysis chain is prohibitively long to be deployed in the experiment. Specifically, preliminary estimations suggest a requirement of 10$\times$ speedup of the track-finding routine. In this context, we explore a GPU-based solution to accelerate track-finding through parallel processing and present the implementation details and the results of our study for different pileup conditions. The results indicate that the GPU solution meets our expectation in terms of execution speed without compromising on the reconstruction efficiency.}


\begin{document}
\maketitle
\flushbottom

\section{Introduction}
The most recent estimate of the muon's magnetic anomaly, \( a_{\mu} \), by the Fermilab Muon \( g{-}2 \) experiment, using data from Runs 1 through 6, has a remarkable precision of 127 parts per billion (ppb), providing a stringent test of the Standard Model (SM)~\cite{muon2025measurement, ParticleDataGroup:2024cfk, PhysRevLett.19.1264, Altarelli:2020jng}. Complementary to the Fermilab experiment, the J-PARC muon \( g{-}2 \)/EDM experiment aims to measure both \(a_\mu\) and the electric dipole moment (EDM) of the muon $d_\mu
$, defined by the following relations:
\begin{equation}
a_\mu = \frac{g-2}{2}, \qquad
\vec{\mu}_\mu = g\!\left(\frac{e}{2m}\right)\vec{s}, \qquad
\vec{d}_\mu = \eta\!\left(\frac{e}{2mc}\right)\vec{s},
\end{equation}
where, \(e\), \(m\), and \(\vec{s}\) are the electric charge, mass, and spin vector of the muon, respectively.
Here, $\mu_{\mu}$ is the magnetic dipole moment of the muon, \(g\),  the Land\'e \(g\)-factor,  and \(\eta\) is a dimensionless quantity related to the EDM~\cite{PhysRevD.97.114025, Jegerlehner:2017gek}. The experimental quantities are determined through a precise measurement of the spin precession frequency of muons stored in a magnetic field~\cite{Thomas01011927}. The  J-PARC muon\textit{g-2}/EDM experiment will demonstrate the measurement using a novel approach with re-accelerated ultra-slow muons of kinetic energy 25 meV instead of the \textit{magic-gamma} technique used by the Fermilab collaboration to independently cross-validate the previous results \cite{10.1093/ptep/ptz030, Grange2018Muong2, KESHAVARZI2022115675, PhysRevD.73.072003}. 

The SM prediction of $a_\mu^{\mathrm{SM}}$ from the 2020 white paper~\cite{AOYAMA20201} and the experimental value of $a_\mu^{\mathrm{exp}}$ from the 2023 Fermilab result~\cite{Aguillard_2023} are as follows, respectively:
\begin{equation}
a_\mu^{\mathrm{SM}} = 116\,591\,810(43)\times 10^{-11}, \qquad
a_\mu^{\mathrm{exp}} = 116\,592\,059(22)\times 10^{-11},
\end{equation}
where the uncertainties are from the electroweak, leading-order hadronic, and higher-order hadronic contributions in $a_\mu^{\mathrm{SM}}$. The errors are the statistical, systematic, and external parameter uncertainties combined in quadrature in $a_\mu^{\mathrm{exp}}$. A comparison between the Fermilab result from Run-1/2/3 presented here, $a_\mu(\mathrm{exp})$, and the prediction from the 2020 white paper yields a discrepancy of $5.0\sigma$. This deviation from the SM prediction could hint at new physics, motivating more precise measurements of \( a_\mu \). 

The result from the current experimental measurement and the updated prediction of \( a_{\mu} \) from white paper 2025 is given by~\cite{muon2025measurement, PhysRevD.98.030001, ALIBERTI20251}:
\begin{equation}
\Delta a_\mu \equiv a_\mu^{\mathrm{exp}} - a_\mu^{\mathrm{SM}} 
= 116592071.5(14.5) \times 10^{-11} \ - 116592033(62) \times 10^{-11} = 38(63) \times 10^{-11}
\end{equation}
Though the new results have reduced the tension between the SM model value and experimental value of \( a_\mu \), it is important and intriguing to validate the experimental result with a new approach and new systematics. 
% The J-PARC muon \textit{g-2}/EDM experiment is designed to measure $a_\mu$ and $\eta$ using a technique different from that used in previous experiments that used the magic gamma approach~\cite{PhysRevD.73.072003}.

With the data-taking expected in 2030 for the J-PARC muon g-2/EDM experiment, simulation studies are being performed to benchmark and understand the performance of the detector and its associated algorithms.
Critical for the success of the experiment is the track reconstruction algorithm which estimates the momentum of the decay positrons and associates them with an unique muon. At this juncture, the track-finding algorithm imposes a performance bottleneck that slows down the track-reconstruction pipeline. Simulation studies have shown that a 10$\times$ speed-up is required to match up with the expected muon rates. To that end, several studies have been undertaken to identify possible approaches to improve the track-finding performance in terms of the execution time and efficiency.

In this study, we explore the use of GPU parallelism to speed up the track-finding process using Geant4-simulated positron tracks generated with the "g2esoft" framework developed for simulation and track reconstruction for the J-PARC muon \textit{g-2}/EDM experiment \cite{owens2008gpu, Agostinelli2003GEANT4}. Section~\ref{Experimental Method} briefly describes the experimental setup of the J-PARC muon $g\!-\!2$/EDM experiment. Simulation studies follow this in section~\ref{Simulation studies}, the track-finding procedure in section~\ref{Track Finding}, and a comparison between CPU- and GPU-based approaches in sections~\ref{CPU-approach} and~\ref{GPU-approach}, respectively. Finally, we analyze the performance results in section~\ref{Analysis} and present our conclusions in section~\ref{Summary and conclusions}.

\section{Experimental method}
\label{Experimental Method}
\begin{figure}[H]
    \centering
    \includegraphics[width=1.0\linewidth]{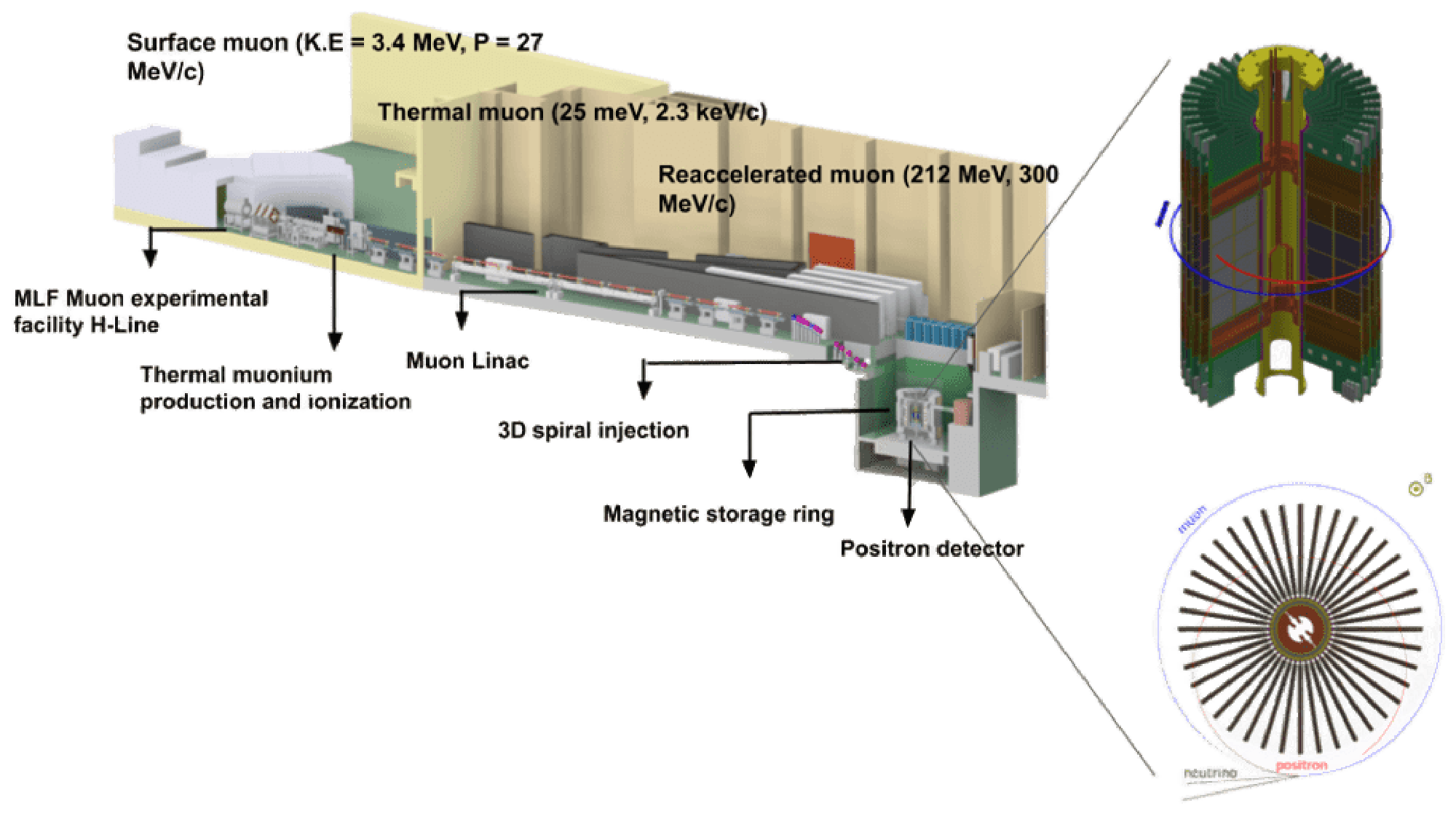}
    \caption{Schematic view of the muon \textit{g - 2}/EDM experiment at J-PARC MLF. Surface muons with a momentum of about 27 MeV/c from the production target start their journey from the H-Line and are brought to rest with about 25 meV kinetic energy through the formation of muonium in a silica aerogel medium. The muons are then transported to the muon LINAC where they are re-accelerated to an energy of 212 MeV to be finally injected into the magnetic storage ring. The muons orbit in the magnetic field until their journey ends as they decay to positrons and neutrinos. These positrons enter the detector volume placed in the storage ring.  Note that positrons with a momentum of about 150 MeV/c cross the central void region of detector creating split tracks. The positron detector volume and the top view with a typical muon orbit and decaying positron orbit are also shown.}
    \label{fig:enter-label22}
\end{figure}

A primary 3\,GeV proton beam from the J-PARC accelerator strikes a graphite target to produce secondary particles, including muons~\cite{muonsource}. Of these, muons which are produced near the surface of the production target and eventually escape, also called surface muons, with  a maximum momentum of 29.8\, MeV/$c$ are extracted through the H-line beamline~\cite{surfacemuons, 10.1093/ptep/pty116}. At a proton beam power of 1\,MW, the expected muon beam intensity reaches approximately $10^8$ muons per second. The surface muons are then injected into a room-temperature silica aerogel target, where they are slowed down, thermalized, and form muonium (Mu) atom~\cite{Tabata:2014lsa, 10.1093/ptep/ptt080}. The thermalized Mu atoms diffuse into an adjacent vacuum region, where they are ionized by laser irradiation to generate ultra-slow muons~\cite{doi:10.7566/JPSCP.21.011060}. At this stage the polarization of the muons is about 50\% which are then re-accelerated to a momentum of 300\,MeV/$c$ and vertically injected into a compact storage ring~\cite{Otani2018, Kondo_2017}. This ring employs a solenoid magnet providing a highly uniform magnetic field of 3 T  with a precision of 1\,ppm within the 66\,cm diameter storage region, made possible through technologies similar to those used in Magnetic Resonance Imaging (MRI) magnets ~\cite{mrimagnet, Iinuma2016}. A novel vertical injection scheme significantly improves injection efficiency by more than an order of magnitude. The ring uses weak magnetic focusing to maintain the muon orbit. The detector volume, based on silicon-strip technology, is placed inside the field to track the decay positrons and to register their hit positions~\cite{inproceedings}. 
% The detector aims to measure the muon's anomalous spin precession frequency, \( \omega_{a} \), and the up-down asymmetry of positron emission caused by the electric dipole moment (EDM). 
After acceleration to 300~MeV/\(c\), the muon beam arrives in 10~ns-wide pulses consisting of three microbunches with a repetition rate of 25~Hz. Each pulse, or "fill", injects about \( 10^4 \) muons into the storage ring. Shortly after each fill, approximately 30 positrons from muon decays are detected within a 5~ns time window~\cite{10.1093/ptep/ptz030}.
Figure~\ref{fig:enter-label22} shows the schematic diagram of the muon \textit{g-2}/EDM experimental setup with graphite target, LINAC, and positron detector (lateral view and top view). The detector consists of 40 radial modules called vanes. Each vane consists of 16 sensors, half of which, called R-sensors, register the radial coordinate and half, called Z-sensors, the axial coordinate of ionization. To ensure good acceptance, the sensitive area along the axial direction is set to $\pm 200~\text{mm}$. For efficient detection of mid momentum positrons ($200~\text{MeV}/c < p < 275~\text{MeV}/c$), the radial sensitive range is from $r = 70~\text{mm}$ to $r = 290~\text{mm}$. The active area of each sensor is $97.28~\text{mm} \times 97.28~\text{mm}$ with a thickness of $0.32~\text{mm}$.
A sensor contains two blocks of 512 strips with a pitch of $190~\mu\text{m}$.
Therefore, a vane has 16{,}384 strips, and the detector has a total of 655k strips.

\section{Simulation studies}
\label{Simulation studies}
\begin{figure}[H]
    \centering
    \includegraphics[width=0.5\linewidth]{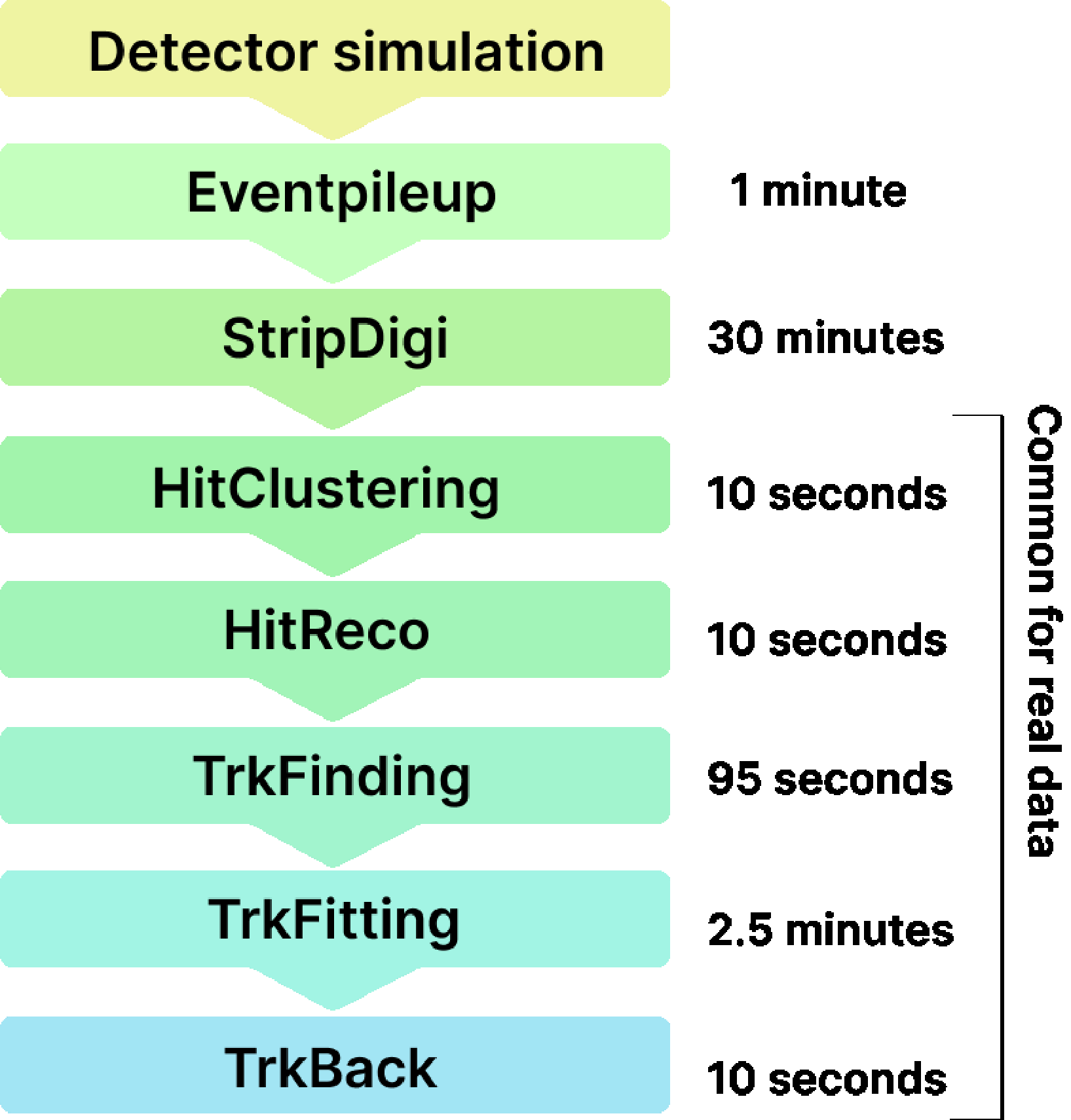}
    \caption{Standard processing order of simulation, digitization and track reconstruction. The typical computation time required for each process is also shown for processing $10^4$ muon decay events in a single CPU.}
    \label{order}
\end{figure}

In order to benchmark the performance of algorithms, a simulation framework for J-PARC muon \textit{g-2}/EDM experiment has been developed. The standard simulation workflow is illustrated in figure~\ref{order}. The first step is the detector simulation stage which uses the Geant4 toolkit to model muon decay and the interactions of decay positrons with detector materials. By default, each event simulates a single muon decay. An ideal 300\,MeV/\emph{c} muon beam circulates at a radial position of \( r \sim 333 \)\,mm and vertical position \( z = 0 \), within a uniform 3\,T magnetic field oriented along the \( z \)-axis. Standard electromagnetic processes are used to model interactions of charged particles with matter, and energy loss due to synchrotron radiation is also included for positrons and electrons.
The next step in simulation, EventPileup, merges multiple events simulated at the detector simulation stage into a single event to emulate the pile-up of muon decay events. The "StripDigi" module performs digitization by converting simulated hits (SimHits) into strip-level hits (StripHits), accurately reconstructing their spatial and temporal information to reflect the detector resolution and response characteristics. The "HitClustering" processor generates "StripClusters" by merging adjacent StripHits. HitReco creates RecoHits by combining StripClusters in R and Z sensors. These "RecoHits" are used as input for track-finding, which is discussed in the subsequent section in a little more detail. Track reconstruction is performed by the "TrkFitting" module, which fits particle trajectories using input data from either "RecoHits" or "StripClusters". These inputs correspond to processed signals from the tracking detectors. The track parameters are estimated at the position of the first detector hit using the Kalman filter algorithm, as implemented in the Genfit2 toolkit~\cite{Bilka2019ImplementationOG}. Once the tracks have been reconstructed, the "TrackBack" module extrapolates them backward to the muon orbit in order to estimate the decay vertex. This step is essential for reconstructing the decay topology and associating tracks with their parent muons. The extrapolation is applied exclusively to tracks that are not identified as ghosts-false or ambiguous reconstructions typically arising from noise or misidentified hits. The positron track is extrapolated starting from the estimated position and momentum at the first detector hit. The muon orbit is modeled as a fixed circular trajectory in a plane perpendicular to the $z$-axis, consistent with the storage ring configuration. In contrast, the positron trajectory is approximated as a helix, arising from its motion in the uniform magnetic field along the $z$-direction. The muon decay position is then estimated as the point of closest approach between the muon and positron trajectories in three-dimensional space. This geometrical construction ensures accurate localization of decay vertices, which is critical for downstream analysis such as lifetime measurements and background suppression.
\section{Track finding}
\label{Track Finding}
\begin{figure}[H]
    \centering
    \includegraphics[width=0.99\linewidth]{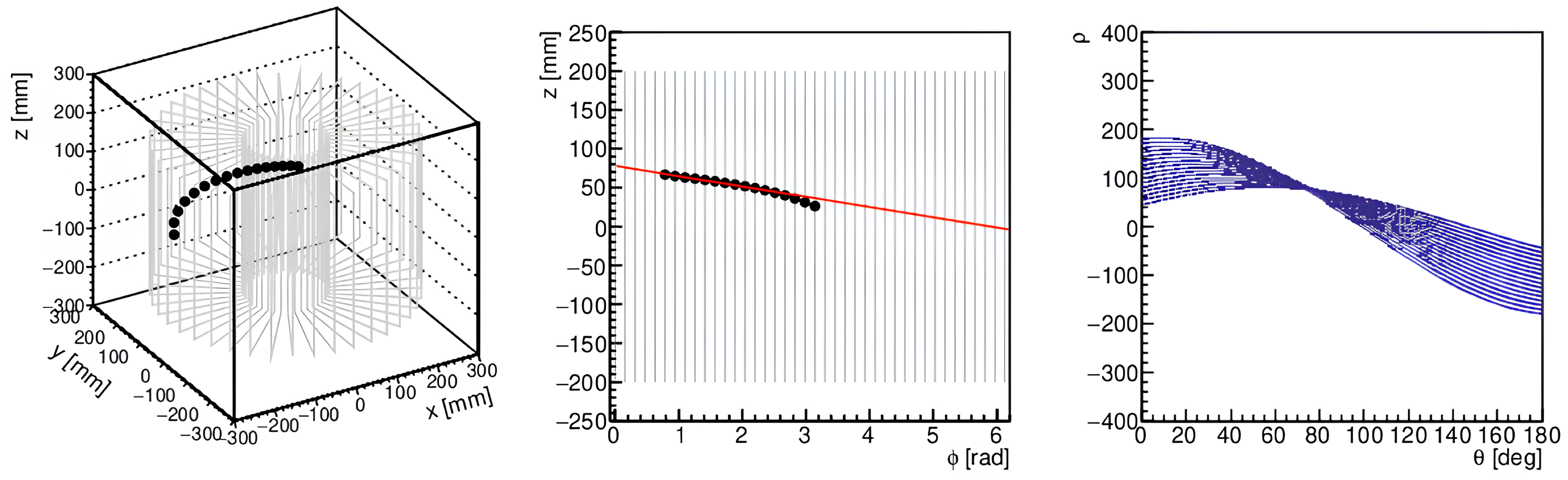}
    \caption{An example muon decay event. Left: RecoHits in three-dimensional space showing the hits due to the positrons. Middle: projection of the track shown in the left onto the $\phi$-$z$ plane. Right: the corresponding Hough space in which each curve is generated from a single hit and the intersection of the curves, i.e, the most populated bin in Hough space corresponds to the red line in the $\phi$-$z$ plane.}
    \label{houghtransform}
\end{figure}

A general example of track-finding for \textit{g-2}/EDM experiment is described in simple steps in figure~\ref{houghtransform}.
Positrons from muon decays travel in a helical trajectory within the magnetic storage ring. When this trajectory is projected onto the $\phi$-$z$ plane, it appears as a straight line with a certain slope and intercept. The straight line model is only an approximation and deviations are expected on account of the fact that the center of the track does not always coincide with the centre of the detector volume in addition to the effect brought about by energy loss and  scattering. To identify these lines, a hough transform algorithm~\cite{osti_4746348, 10.1145/367177.367199} is employed. This technique is particularly effective in detecting geometric shapes such as lines or circles even in data that is noisy, sparse, or has overlapping patterns.
Each point in the $\phi$-$z$ plane can be transformed into a sinusoidal curve in the Hough space, defined by the parameters $(\rho, \theta)$, through the relation~\cite{10.1145/361237.361242}:
\begin{equation}
    \label{eq:ht}
    \rho = \phi \cos\theta + z \sin\theta
\end{equation}
In the Hough space, intersections of these curves represent potential straight lines formed by multiple hits in the $\phi$-$z$ plane. The bin with the highest accumulation in Hough space indicates the most probable line candidate. Once a line is identified in the $\phi$-$z$ plane, the corresponding RecoHits lying along it are used as seeds to reconstruct the full track in three-dimensional space. 

Track extension is performed by selecting three RecoHits (track seed) on consecutive vanes and extrapolating their trajectory to search for additional hits on neighboring vanes in the forward or backward direction. This search is confined to fixed time window of width 10 ns. When no more RecoHits can be added within the current window, the window is shifted by a time step of 5 ns, and a new track-finding process begins. If the flag for finding the tracks in the next time window is set to \textit{"true"} from the previous time window, then the algorithm first attempts to continue building the previously identified track candidates in the new time window before initiating a completely new search. This iterative procedure ensures efficient reconstruction of positron tracks even in high-occupancy or pile-up scenarios.

\section{CPU-based approach}
\label{CPU-approach}
Track-finding on the CPU is performed sequentially through the following steps. First, the reconstructed hit points are divided into overlapping time windows of 10~ns, with a sliding step of 5~ns. Within each time window, the spatial coordinates \((x, y, z)\) of the hits are transformed into the \((\phi, z)\) space, where \(\phi = \tan^{-1}(y/x)\). The transformed \((\phi, z)\) points are then mapped to a Hough space using the hough transform equation (Eq.~\ref{eq:ht}). This results in a 2D histogram in the \(\theta\)-\(\rho\) space with 180 bins in \(\theta\) (ranging from \(0^\circ\) to \(180^\circ\)) and 1000 bins in \(\rho\) (ranging from \(-500\) to \(500\)). Each hit contributes to multiple bins based on its possible track parameters. After the histogram is filled, a peak-finding algorithm identifies the bin with the highest number of entries. The corresponding \((\theta, \rho)\) values are interpreted as the most probable track parameters in the \((\phi, z)\) plane. Using these peak values, track seeds are found by selecting the hits that align with the straight line formed by the identified \((\theta, \rho)\). These seeds serve as the initial candidates for tracks. A track extrapolation step then follows, which refines the track by extending it across vanes and associating additional hits based on geometric and timing constraints.
This entire process—time slicing, coordinate transformation, Hough mapping, peak detection, seed finding, and track extrapolation—is repeated for each time window. While conceptually straightforward, this sequential CPU-based approach is computationally intensive, especially in sorting hits, populating the Hough histogram, and extrapolation.\\
Figure~\ref{order} shows the sequence of simulation steps along with the computation time for processing $10^4$ muon decay events in a single CPU. All processes beginning from \textit{HitClustering} to \textit{TrkBack} are relevant for the experimental data while the others are required only for simulated data. The most time-consuming parts are track- finding and track-fitting, which currently take approximately 95 seconds and 280 seconds, respectively, for $10^4$ muon decays.

Currently, the data processing rate is approximately $10^4$ muons every 7 minutes, which corresponds to  24 muons per second per CPU. 
With 1000 CPUs, the overall data processing rate reaches roughly $2.4 \times 10^4$ muons per second. 
However, the experiment is expected collect around $10^{13}$ muon decays in total. 
At a beam intensity of $10^4$ muons per pulse at 25~Hz, yielding a rate of $2.5 \times10^5$ muons per second, the data processing pipeline is also expected to run at the same rate \cite{10.1093/ptep/ptz030}.
% The beam intensity will be $4 \times 10^4$ muons per pulse at 25~Hz, yielding a rate of $10^6$ muons per second. 
% Therefore, to keep up with the data taking rate, the software must be capable of processing data from approximately $10^6$ muon decays per second. 
This implies that the current software speed needs to improve by a factor of about 10 to meet the experimental requirements.

\section{GPU-based approach}
\label{GPU-approach}
Unlike CPUs, which have only a few powerful cores, GPUs have a much larger number of threads (cores) that can run at the same time. While each individual thread may be slower than a CPU core in terms of processing power, the massive parallelism allows GPUs to handle large-scale computations efficiently. In the J-PARC muon $g\!-\!2$/EDM experiment, track- finding involves processing a large number of detector hits to reconstruct particle trajectories. Traditionally, this process is performed in sequence on a CPU, starting from the hough transform, followed by helical track seed fitting, and then extrapolation. This sequential processing is computationally expensive due to the high volume of hit data.
To improve performance, we utilize a GPU-based approach, where detector hits are processed in parallel. In this method, each detector hit is mapped to a GPU thread, and a certain number of threads form a block. All threads within a block work together to fill a histogram in shared memory (as described in later in this Section), allowing efficient collection of votes during the hough transform step. Track seed finding, helical-track fitting, and extrapolation are also performed in parallel across threads for each block. To organize the processing, hits are divided into 10~ns time windows, with each GPU block handling one time window. Within each block, all steps—from hough transform to track fitting—are carried out simultaneously, significantly speeding up the reconstruction. 
% For example, assuming processing of one time window on the CPU takes $t$ seconds. Then, processing $100$ such windows sequentially would take; $\sim$ 100 $\times$ $t \quad 
% \text{seconds}$. 
While individual GPU cores are slower than CPU cores, since the threads run concurrently, the execution time of the entire job is, in the best-case scenario, roughly the time taken by a single thread. 

% If processing one time window on the GPU takes $10\timest$ seconds due to thread overheads, all $100$ windows can still be processed simultaneously in; $\sim$ 10$\times$ $t \quad \text{seconds}$. 

\begin{figure}[H]
    \centering
    \begin{subfigure}[b]{0.45\textwidth}
        \centering
        \includegraphics[width=\textwidth]{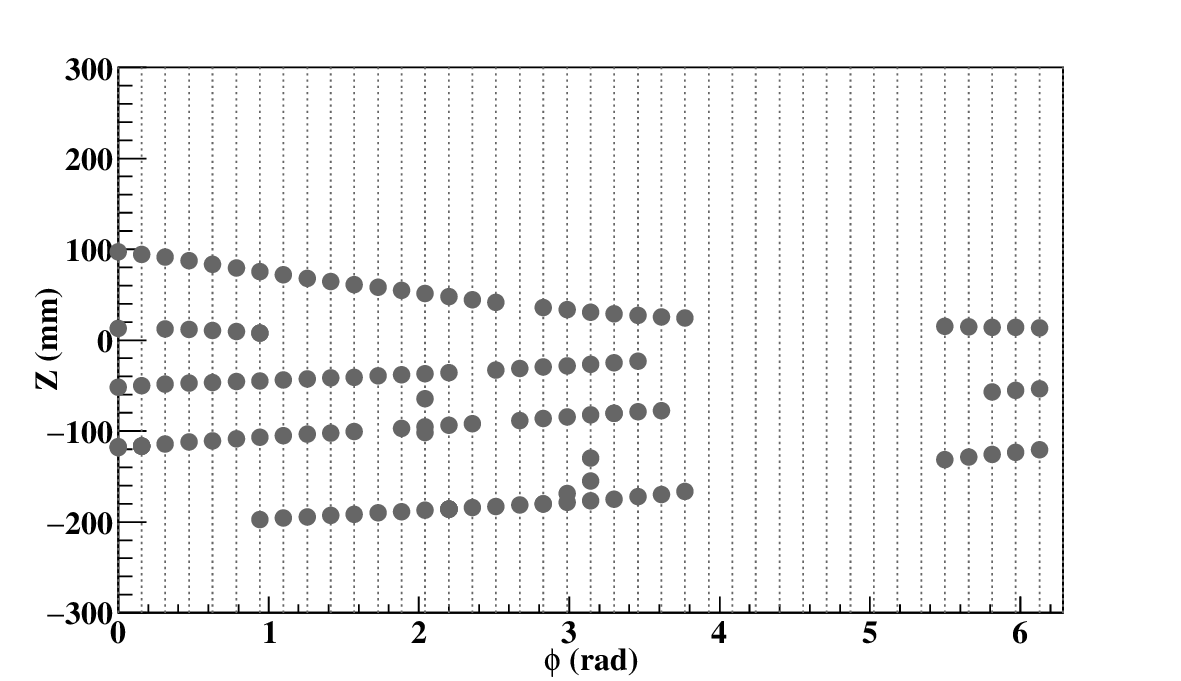}
        \caption{Hits in the $\phi$-z plane}
    \end{subfigure}
    \begin{subfigure}[b]{0.45\textwidth}
        \centering
        \includegraphics[width=\textwidth]{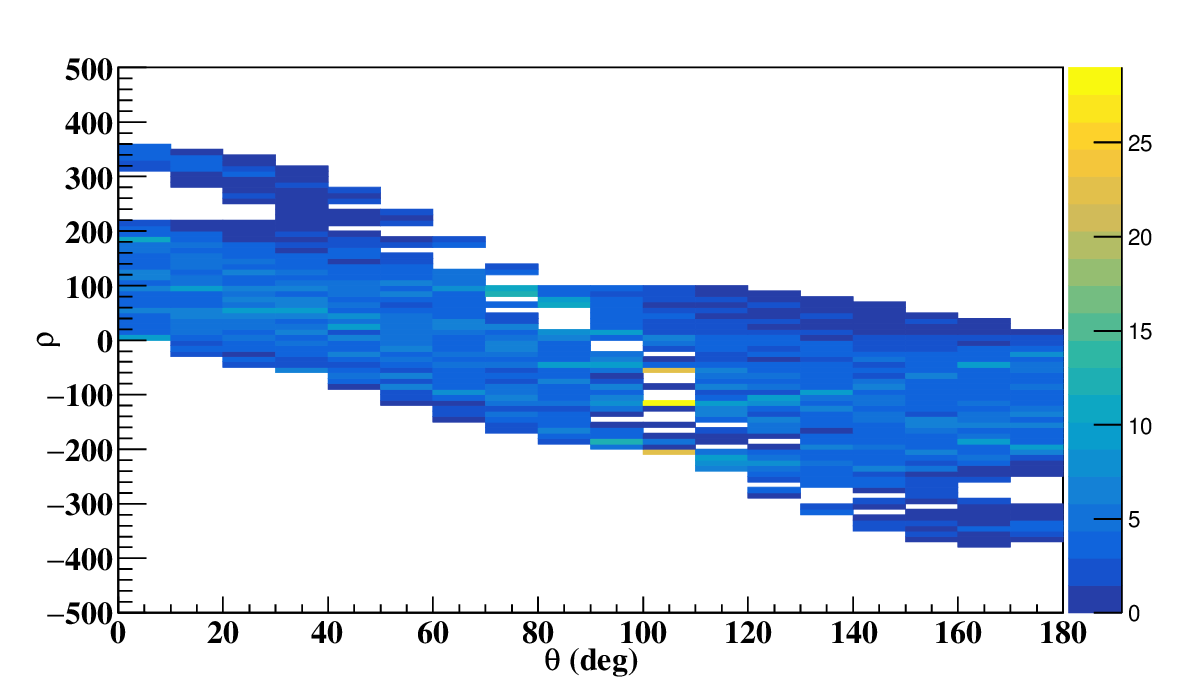}
        \caption{Coarse bin histogram in Hough space}
    \end{subfigure}
    
    \begin{subfigure}[b]{0.45\textwidth}
        \centering
        \includegraphics[width=\textwidth]{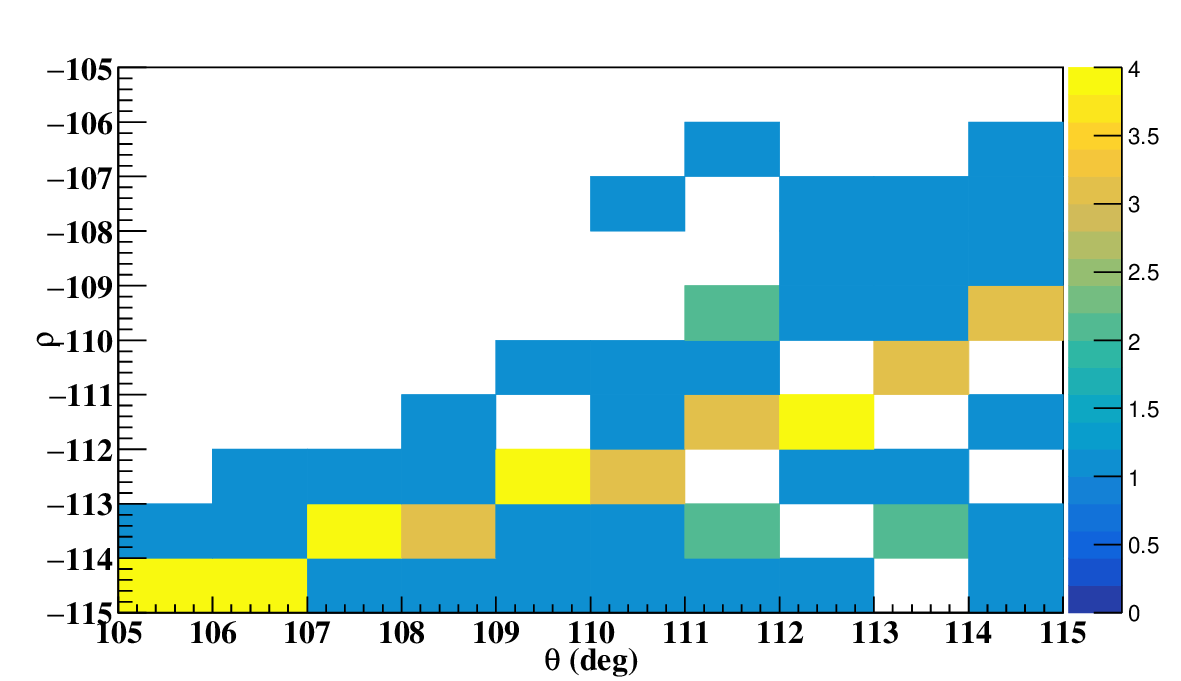}
        \caption{Fine bin histogram in Hough space}
    \end{subfigure}
    \begin{subfigure}[b]{0.45\textwidth}
        \centering
        \includegraphics[width=\textwidth]{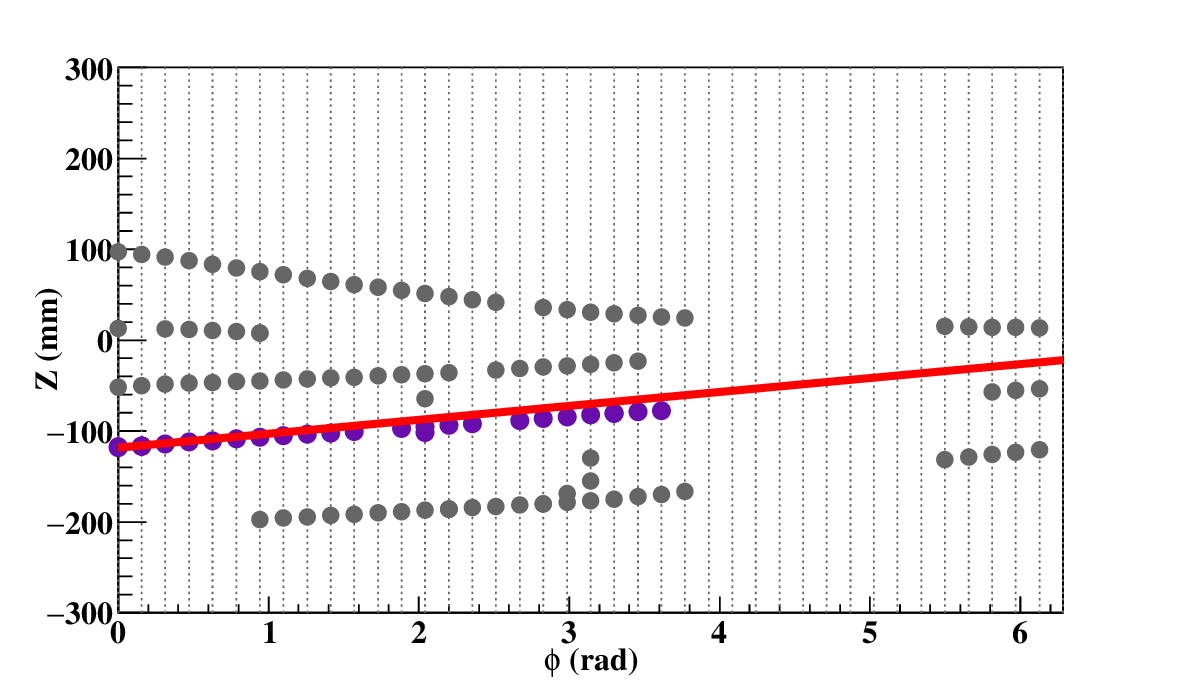}
        \caption{Identified track seed}
    \end{subfigure}
    
    \caption{Steps in the hough transform for track seed identification, showing hits in the $\phi$–$z$ plane, Hough space histograms, and the final seed. (a) is the projection of positron tracks of a multi-muon decay (pileup) event in the $\phi-z$ plane (b) is the same information transformed to the hough-space with a coarse binning of 18 bins along $\theta$ and 100 bins along $\rho$ with the yellow bin regions having a high likelihood of a track. (c) is a fine binned version (10 bins along each axis) of  (b) around the high likelihood zone identified in (b). (d) is the line obtained from the bin with the highest count in (c).}
    \label{fig:four_photos}
\end{figure}

Thus the GPU-based approach, by leveraging massive parallelism, achieves a significant speed-up in track reconstruction despite slower individual cores.
As shown in figure~\ref{order} and discussed in the previous section, the total time required to process one event ($10^4$ muon tracks) of experimental data is approximately 420 seconds. The most time-consuming steps are track-finding and track-fitting. To achieve our goal of reducing the processing time to around 40 seconds, the overall computation time must be improved by a factor of 10, as noted in section~\ref{CPU-approach}. All the steps mentioned in section \ref{CPU-approach} are done in sequence on the CPU which make them computationally expensive making GPUs a good alternative. 
% At this point GPU can be a suitable alternative to achieve acceleration in the track-finding process. A GPU (Graphics Processing Unit) is a processor designed to handle many calculations at the same time using thousands of parallel cores. 
In track-finding, the GPU speeds up the process by performing the coordinate transformation, Hough mapping, and track reconstruction in parallel.\\
GPU memory differs from CPU memory in both structure and access patterns~\cite{cuda_mem_model}. Each GPU block has its own \textit{shared memory}, which is accessible by all threads within that block. In addition to shared memory, GPUs also include several other types of memory, each designed for specific use cases. However, the amount of shared memory available per block is limited. This poses challenges when constructing large data structures like the Hough space. To overcome this, we perform the hough transformation into two stages: a coarse-binning histogram followed by a fine-binning histogram. This approach allows us to meet the binning requirements specified in the experiment while staying within the GPU's shared memory constraints, as illustrated in figure~\ref{fig:four_photos}.

% \begin{figure}[H]
% \centering
% \includegraphics[width=0.7\linewidth]{Group_4.eps}
% \caption{Schematic overview of the GPU memory distribution and its components~\cite{cuda_mem_model}.}
% \label{fig:memory}
% \end{figure}

\section{Analysis}
\label{Analysis}
\begin{figure}[H]  % or [htbp] if you prefer LaTeX to float the figure
    \centering
    \begin{subfigure}[t]{0.48\textwidth}
        \centering
        \includegraphics[width=1.15\linewidth]{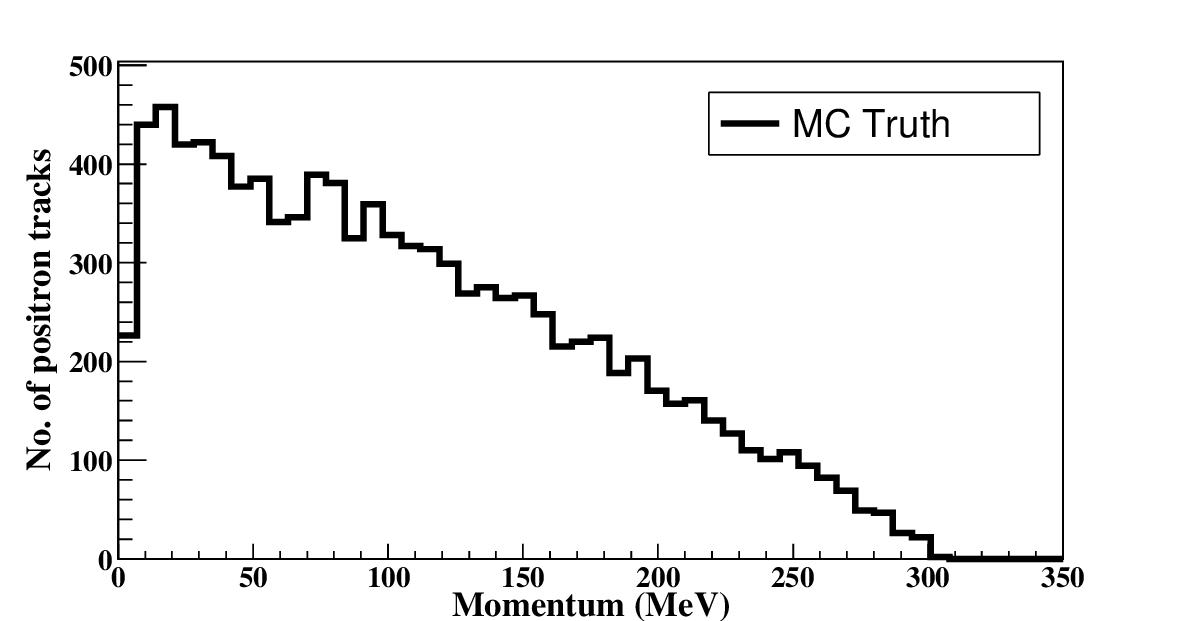}
        \caption{}
        \label{fig:image1}
    \end{subfigure}
    \hfill
    \begin{subfigure}[t]{0.48\textwidth}
        \centering
        \includegraphics[width=1.15\linewidth]{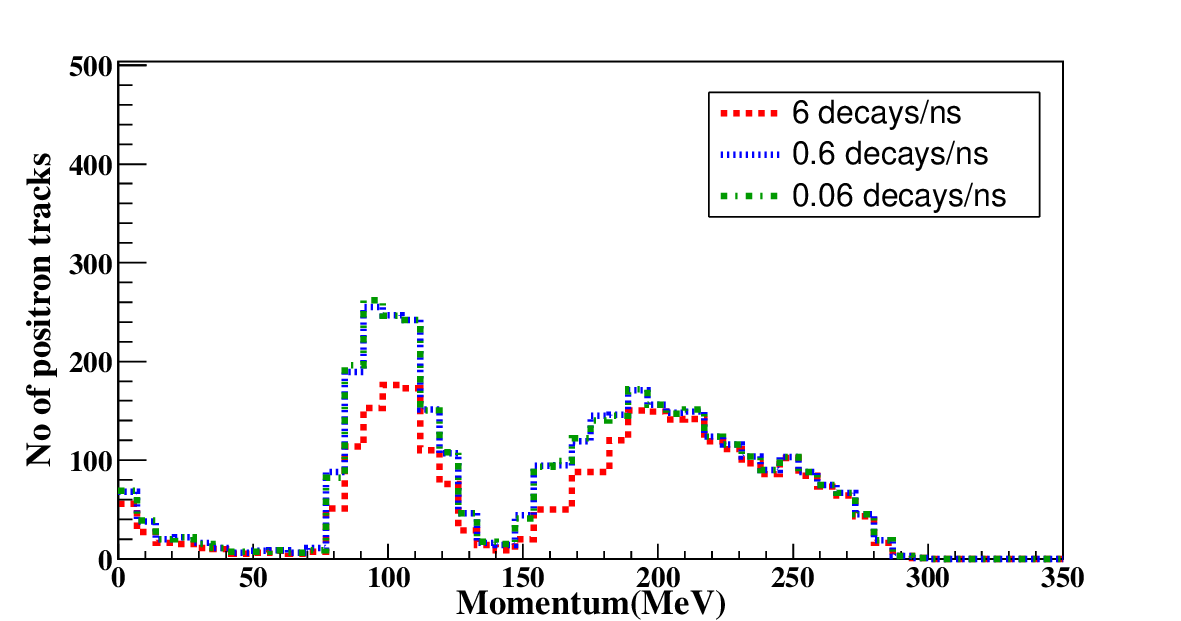}
        \caption{}
        \label{fig:image2}
    \end{subfigure}

    \caption{(a) Generator level momentum distribution of decay positrons from Geant4 (b) The momentum distribution from decay positron tracks  identified by the track-finding algorithm.}
    \label{fig:side_by_side}
\end{figure}

Using the simulator and the track-finding tool, the track-finding efficiency in a multiple-track condition is estimated. The efficiency, $\varepsilon$, is defined as follows:

\begin{equation}
\varepsilon = \frac{N_{\text{found}}}{N_{\text{all}}},
\end{equation}

\noindent where $N_{\text{found}}$ and $N_{\text{all}}$ are the numbers of found tracks and all tracks, respectively. A track is regarded as found if its hits are reconstructed and clustered in continuous five vanes of the detector. We note that tracks with momentum greater than 200~MeV/$c$ have a sufficient number of hit points in the silicon vanes.\\
The generated momentum distribution and the reconstructed momentum distribution for different pileup rates of the positron tracks are shown in figure \ref{fig:side_by_side} respectively. The plot is marked by lower efficiencies below about 80 MeV/c due to tracks without sufficient hits for reconstruction and a significant dip at about 150 Mev/c due to tracks that are disjoint as they cross the void central volume of the detector (refer figure \ref{fig:enter-label22}).

\begin{figure}[H]
    \centering
    \includegraphics[width=0.8\linewidth]{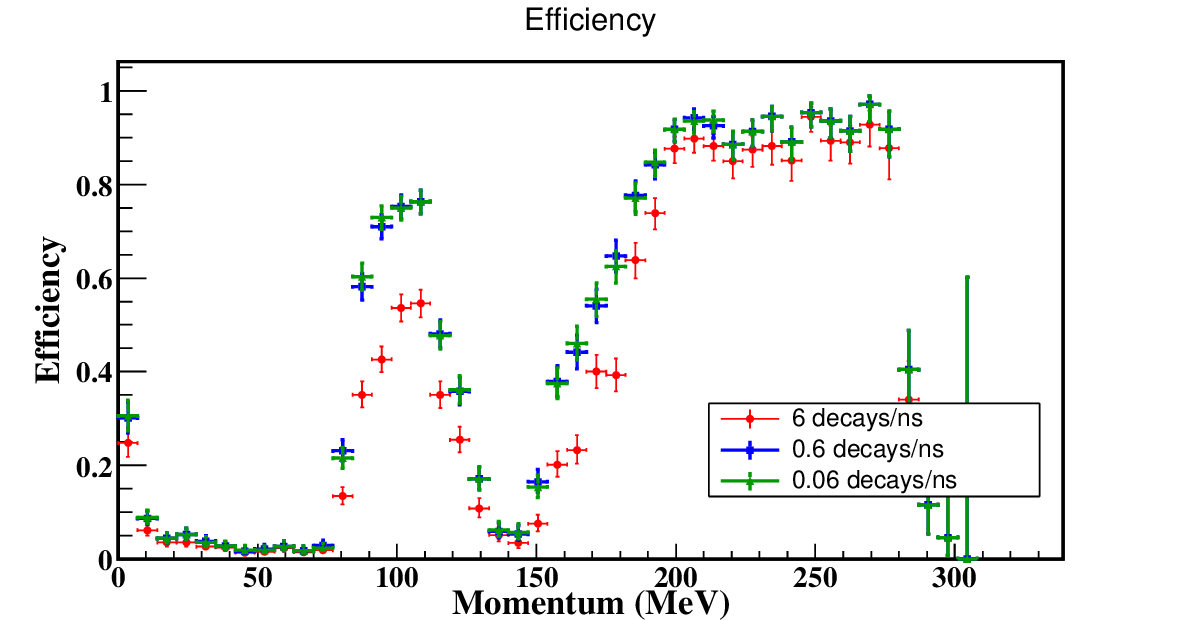}
    \caption{Momentum dependence of track-finding efficiency with different conditions of track rates from 0.06 tracks/ns to 6 tracks/ns.}
    \label{fig:c3}
\end{figure}

Figure \ref{fig:c3} shows the track-finding efficiency for a single track as a function of the positron momentum under different track rate conditions. At the currently proposed intensity of $10^4$ muons/pulse, equivalent to $2\times10^5$ muons per second leading to an estimated highest decay rate of 1.5 decays/ns.  The study shows that even for a much higher track rate of 6 tracks/ns, the finding efficiency still remains above 90\% in the momentum range of $200~\text{MeV}/c < p < 275~\text{MeV}/c$, the window that is also useful for muon g-2 analysis.

Tracking detectors can occasionally reconstruct false hits and tracks, known as ghost hits. Ghost hits are incorrect combinations of R-sensors hits and Z-sensors hits used in track reconstruction which can impact the performance of the track reconstruction. To evaluate their impact, the contribution of ghost hits in reconstructed tracks is studied for single-track events. It is found that approximately 50\% of the tracks are found without any ghost hits, and the majority of the remaining tracks contain less than 20\% ghost hit contribution. Some of the contributions from ghost hits are removed during the track-fitting process, and with a more optimized track-finding algorithm, they can be further reduced.
\begin{figure}[H]
    \centering
    \includegraphics[width=0.7\linewidth]{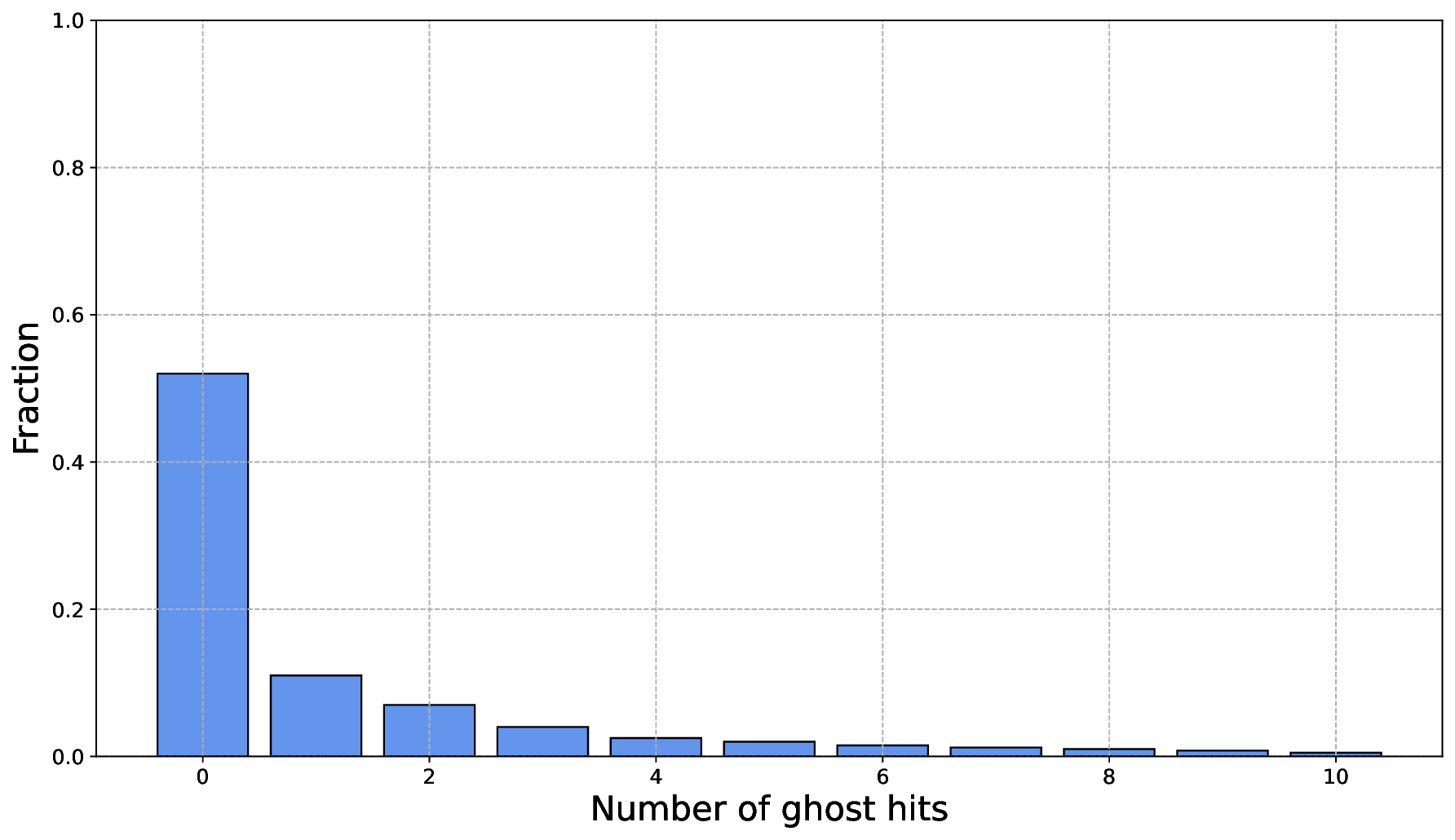}
    \caption{The fraction of ghost hits identified  by the track-finding algorithm.}
    \label{fig:ght}
\end{figure}

\begin{figure}[H]
    \centering
    \includegraphics[width=0.7\linewidth]{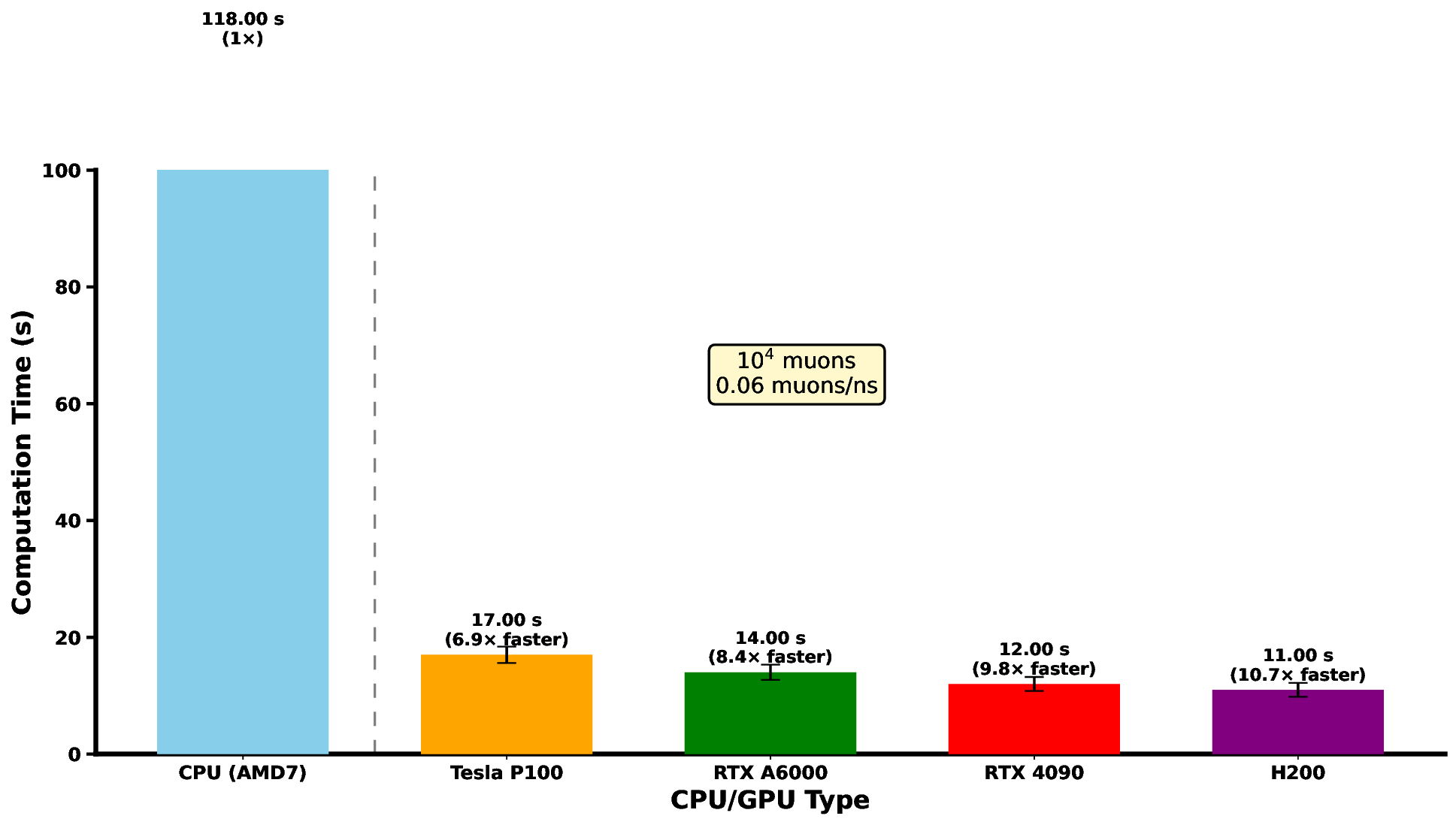}
    \caption{Approximated GPU performance comparison with different GPU architectures for 0.06 muon decays/ns}
    \label{fig:performance}
\end{figure}

\indent The computation times for GPU activities, including kernel execution and data transfers between host and device (both Host-to-Device and Device-to-Host), are summarized in table~\ref{tab:example-taller}. These measurements were obtained using the \texttt{nvprof} profiling tool for track rates ranging from 0.06 to 6 tracks/ns on a Tesla P100-PCIE GPU with 16~GiB of memory, using $10^4$ simulated muon decay events ~\cite{nvprof}.
\begin{table}[H]
\renewcommand{\arraystretch}{1.7} % Increase row height
\centering
\begin{tabular}{|c|c|c|c|c|}
\hline
\textbf &  & \textbf{0.06 $\mu$ decay/ns} & \textbf{0.6 $\mu$ decay/ns} & \textbf{6.0 $\mu$ decay/ns} \\
\hline
   \textbf{GPU based} & GPU Kernel & 17.00 s & 58.00 s & 118.00 s \\
\hline
   & Memcpy HtD & 70.00 ms & 70.00 ms & 60.00 ms\\
\hline
   & Memcpy DtH & 90.00 ms & 90.00 ms & 80.00 ms\\
\hline
  \textbf{CPU based} & CPU (AMD7) & 119 s & 240 s & 345 s\\
\hline
\end{tabular}
\caption{Approximate computation time for GPU kernel execution, CPU, and memory transfer rates between host to device and device to host, at varying muon decay rates of $10^4$ muons, measured using \texttt{nvprof}, a GPU performance profiling application ~\cite{nvprof}.}
\label{tab:example-taller}
\end{table}
The computation time on GPUs depends on several key factors. These include the memory bandwidth, compute capability, number of processing cores, and clock speed. Other important factors are the amount of shared memory, register size, and how efficiently the algorithm uses memory. In addition, performance is influenced by the number of threads that can run in parallel (occupancy), the overhead of launching GPU kernels, and the time taken to transfer data between the CPU and GPU. Table~\ref{tab:gpu_comparison} compares different GPUs based on important features like architecture, bandwidth, CUDA cores, and performance. A comparison is done on performance between different GPUs currently available in the market with $10^4$ muons decaying at the rate of 0.06 muons/ns. Figure \ref{fig:performance} shows the performance comparison between different GPUs. 

\begin{table}[H]
\centering
\renewcommand{\arraystretch}{1.3}
\begin{tabular}{|l|c|c|c|c|}
\hline
\textbf{Spec / GPU} & \textbf{Tesla P100} & \textbf{RTX A6000} & \textbf{RTX 4090} & \textbf{NVIDIA H200} \\
\hline
Architecture    & Pascal        & Ampere        & Ada Lovelace   & Hopper \\
\hline
FP32 TFLOPS     & 10.6          & 38.7          & 82.6           & $\sim$60 \\
\hline
Tensor TFLOPS   & 21.2          & 312           & $\sim$330      & 2,000+ \\
\hline
Memory Size     & 16 GB HBM2    & 48 GB GDDR6   & 24 GB GDDR6X   & 141 GB HBM3e \\
\hline
Bandwidth       & 732 GB/s      & 768 GB/s      & 1,008 GB/s     & $\sim$4,800 GB/s \\
\hline
CUDA Cores      & 3,584         & 10,752        & 16,384         & 132 SMs + Tensor Cores \\
\hline
\end{tabular}
\caption{Architecture comparison of Tesla P100, RTX A6000, RTX 4090, and NVIDIA H200 GPUs~\cite{nvidiaP100,nvidiaA6000,nvidia4090,nvidiaH200}}.
\label{tab:gpu_comparison}
\end{table}
It is clear at this point that the implementation of the track-finding on a GPU can lead to a speedup improvement between 7-11 $\times$ compared to the CPU version depending on the GPU architecture.
\section{Summary and conclusions}
\label{Summary and conclusions}
Muon \textit{g-2}/EDM experiment at J-PARC measures the $a_{\mu}$ and EDM with a new approach to reach the uncertainty of 460 ppb. One of the crucial part of the analysis lies in the reconstruction of the positron tracks. These positron tracks are reconstructed from the hits in silicon strip detector. 
In the experiment, with an expected muon rate of $10^6$ muons per second, the data collected should also be ensured to be processed at at the same speed. The CPU-based track-finding algorithm, identified to a performance bottleneck, required a 10$\times$ speedup to commensurate with the data collection rate.
% In the experiment, data collection (acquisition) can run the expected speed of $10^4$ muons every 7 minutes (24 muons/sec/CPU). With 1000 CPUs, the data processing rate reaches about $10^5$ muons per second. 
% The software must handle data from approximately $10^6$ muon decays per second (see section~\ref{CPU-Based Approach}).\\

% \indent

With the implementation of a GPU-based algorithm for track-finding, we are able to accelerate the computation time by a factor of around 7 - 11 $\times$ for the pileup rate of 0.06 muon decay/ns (see table \ref{tab:example-taller}), depending on the architecture. This is a forward-looking development expecting that GPUs will be much more commonly used in near future. The efficiency for different pileup rates (0.06 tracks/ns, 0.6 tracks/ns and, 6 tracks/ns) is estimated to be 90-95\%, matching well with the expectations from the CPU-based algorithm. We evaluated the ghost hit contamination at the lowest pile-up rate. As shown in figure~\ref{fig:ght}, approximately 50\% of the tracks are reconstructed without any ghost hits. Among the remaining tracks, the majority exhibit a ghost hit contribution of less than 20\%. As discussed in section~\ref{Analysis} and as inferred from figure~\ref{fig:performance}, GPU-based processing significantly speeds up the track-finding process.

This improvement is attributed to the GPU’s higher compute capability, greater memory bandwidth, and large number of parallel processing cores, offering a clear advantage over traditional CPU-based methods. In particular, the  H200 achieves about 11 $\times$ improvement in performance compared to CPU-based methods.

In the simulation framework, track fitting is one of the most time-consuming steps. It uses input from either \texttt{RecoHits} or \texttt{StripClusters}, which are processed signals from the tracking detectors. The track parameters are estimated at the position of the first detector hit using the Kalman filter algorithm, as implemented in the Genfit2 toolkit~\cite{Bilka2019ImplementationOG}. To further speed up the simulation, the next step in our study could be to implement the track fitting algorithm on the GPU. By integrating both the track-finding and track-fitting algorithms to run entirely on the GPU, we expect to significantly accelerate the overall simulation process.

\section*{Acknowledgements}
This work was supported in part by JSPS Kakenhi Grants No. 20H05625 and 22K21350.
%Bibliography

%\bibliographystyle{unsrt} % or JHEP/unsrtnat etc.
\bibliographystyle{JHEP}
\bibliography{biblo}
\end{document}